41# Principle-driven Fiber Transmission Model based on PINN Neural Network

">Yubin Zang, Zhenming Yu, Kun Xu, Xingzeng Lan, Minghua Chen, Sigang Yang and Hongwei Chen[*]abstract">*Abstract* — In this paper, a novel principle-driven fiber transmission model based on physical induced neural network (PINN) is proposed. Unlike data-driven models which regard fiber transmission problem as data regression tasks, this model views it as an equation solving problem. Instead of adopting input signals and output signals which are calculated by SSFM algorithm in advance before training, this principle-driven PINN based fiber model adopts frames of time and distance as its inputs and the corresponding real and imaginary parts of NLSE solutions as its outputs. By taking into account of pulses and signals before transmission as initial conditions and fiber physical principles as NLSE in the design of loss functions, this model will progressively learn the transmission rules. Therefore, it can be effectively trained without the data labels, referred as the pre-calculated signals after transmission in data-driven models. Due to this advantage, SSFM algorithm is no longer needed before the training of principle-driven fiber model which can save considerable time consumption. Through numerical demonstration, the results show that this principle-driven PINN based fiber model can handle the prediction tasks of pulse evolution, signal transmission and fiber birefringence for different transmission parameters of fiber telecommunications.

*Index Terms*—Fiber Optics; Neural Networks; Principle-driven## I. Introduction

Fiber transmission models provide the effective numerical information for predicting fiber optics or telecommunications. Nonlinear Schrödinger Equation (NLSE) is the basic mathematical model to describe propagation mechanisms of the fiber [1]. The most widely used numerical algorithm which can solve NLSE is split step Fourier method (SSFM) [2-3]. However, SSFM suffers from high computation burden due to iteration steps. To avoid this problem, data-driven models with high time efficiency have attracted research attentions in recent years [4-5]. These models take advantages of the latest achievements of artificial intelligence and apply several neural network structures such as BiLSTM [4] and GAN [5] in the field of fiber transmission [6-10]. By regarding the fiber transmission link as a totally black box, these models block any physical principles behind fiber transmission systems and learn the rules of fiber transmission only through large amount of labeled data which were calculated by SSFM in advance before training. After that, these models can predict signals after fiber transmission in much shorter time. However, the characteristic that the large amounts of labeled data for training must be prepared by SSFM and processed in advance before the training starts can cause extra time consumption in data preparations. This characteristic will also cause data-driven models fail to be trained under some circumstances where the desired signals after transmission can hardly be obtained. Besides, since data-driven models block any physical principles and only learn from labeled training data, their performances are relatively closely related with the distribution of training data which means that these models can hardly perform well in data whose distribution is slightly different from those they processed during training. This explains both BiLSTM [4] and GAN [5] based fiber transmission model can barely predict PAM signals if they were trained with OOK signals. Moreover, most data-driven models barely can provide how signals change all along the fiber propagation in one calculation since each sample of their dataset only consists of signals before and after the fiber transmission.

In order to address the above problems, we propose the principle-driven PINN [11-12] based fiber transmission model in this paper. Thanks to the principle-driven scenario it adopts, this model can be effectively trained without data labels which are originated from signals after fiber transmission and prepared by SSFM through introducing physical meaning in both training and testing. Instead of regarding fiber transmission as a pure mathematical regression issue which trains the models directly through large amounts of data, this model takes fiber transmission principles into account and solves NLSE problem by neural networks. By containing both the initial conditions of pulses or signals before fiber transmission and NLSE into its loss functions, it can progressively learn the transmission rules. Without losing generality, effects including attenuation, fiber dispersion, non-linearity and etc. are taken into consideration in the model. After the model is appropriately trained, in total three different kinds of tasks consisting of pulse evolution, signal transmission and fiber birefringence are conducted to test the performances of the model in fields of both optical telecommunications and fiber optics.

Overall, this paper will be divided into four main parts. All brief background of fiber transmission models and motivations of establishing our principle-driven PINN based model are firstly introduced in the introduction part. Then model structures and configurations will be illustrated in detail in the second part. Since most of the model hyper-parameters'

">Yubin Zang, Xingzeng Lan, Minghua Chen, Sigang Yang and Hongwei Chen are with the Beijing National Research Center for Information Science and Technology (BNRist) and Department of Electronic Engineering, Tsinghua University, Beijing, 100084, China.
Zhenming Yu and Kun Xu are with the State Key Laboratory of Information Phonetics and Optical Communications, Beijing University of Post and Telecommunications, Beijing, 100876, China.
* Corresponding author: chenhw@tsinghua.edu.cn

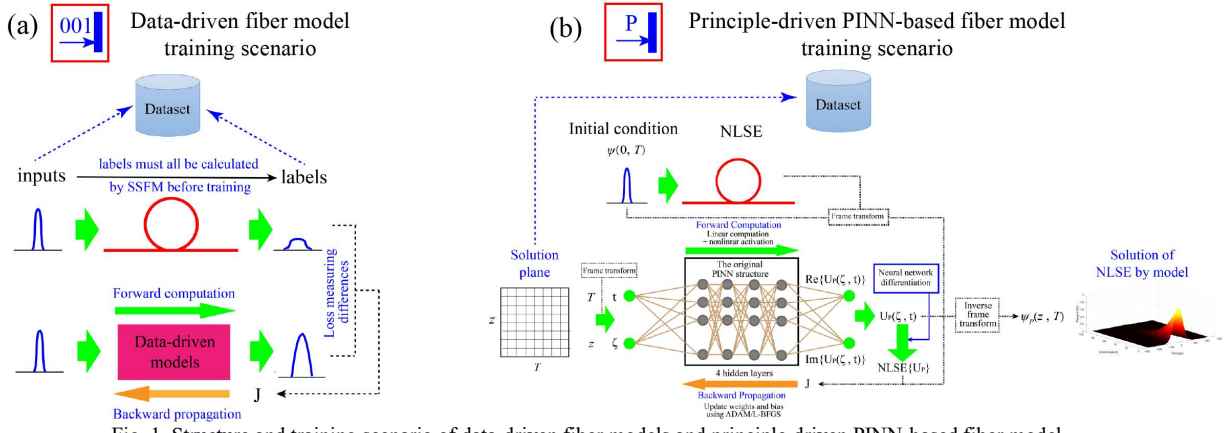

Fig. 1. Structure and training scenario of data-driven fiber models and principle-driven PINN-based fiber model.

TABLE I
IMPORTANT PARAMETERS OF FIBER MODEL FOR THREE DIFFERENT TASKS

| Parameter | Task I | Task II | Task III |
|---|---|---|---|
| $\alpha$ | $4.605\times10^{-5}$ /m | $4.605\times10^{-5}$ /m | $4.605\times10^{-5}$ /m |
| $\lambda_0$ | $1.55\times10^{-6}$ m | $1.55\times10^{-6}$ m | $1.55\times10^{-6}$ m |
| $D$ | 15.6916 ps/nm/km | 17 ps/nm/km | 17 ps/nm/km |
| $S$ | -0.12332 ps/nm$^2$/km | 0.056 ps/nm$^2$/km | 0 |
| $\gamma$ | 0.0013 W$^{-1}$/m | 0.0013 W$^{-1}$/m | 0.0013 W$^{-1}$/m |
| $\tau$ | $2.6\times10^{-15}$ s | $2.6\times10^{-15}$ s | 0 |
| $A_{eff}$ | $8\times10^{-11}$ m$^2$ | $8\times10^{-11}$ m$^2$ | $8\times10^{-11}$ m$^2$ |

determinations are closely related with different tasks, all configurations of pulse evolution, signal transmission and fiber birefringence will be described respectively. Training and testing configurations of the model including optimizer and loss function settings will be illustrated in this part as well. For the third part, performing results will be shown and analyzed in detail for each of the three tasks. Discussions related with the results will then be developed if necessary. Conclusions and future research work will be illustrated in the last part.

## II. MODEL STRUCTURES AND CONFIGURATIONS

### A. Model structures and general configurations

Both the structures and the corresponding training scenarios of previously proposed data-driven models and principle-driven PINN based fiber model is depicted in Fig.1. It can be vividly seen that from the overall perspectives of the model, data-driven models regard fiber transmission as a pure regression problem while principle-driven PINN based model views it as a NLSE solving problem. Therefore, both inputs and outputs of data-driven models should be signals before and after fiber transmission while inputs and outputs of the principle-driven PINN based fiber model should be time-distance frame values and the real and imaginary parts of NLSE solutions. This difference in overall perspectives on fiber transmission also causes different training scenarios. Data-driven models focus on minimize the differences between labels and their outputs while our principle-driven model tries its best to let its outputs satisfy both initial conditions and NLSE equations. As a result, loss functions of data-driven models must at least contain model outputs and labels as is shown in dashed line Fig.1(a) while loss functions of the principle-driven PINN based fiber model should at least measure the extent of wellness about how the outputs from the principle-driven PINN based fiber transmission model obeys NLSE and initial conditions. This indicates that both inputs and labels which are originated from signals before and after transmission respectively are ought to be contained in datasets of data-driven models while no labels are needed in the dataset of the principle-driven PINN based fiber model as can be indicated from the composition of dataset in both models from Fig.1.

Core structure of the principle-driven PINN based fiber transmission model is based on the original PINN structure proposed by Maziar Raissi *et al*. As is shown Fig.1(b), the basic structure is a multi-layer fully-connected neural network. There are two neurons in both input and output layer and 100 neurons for each of the several hidden layers [11-12]. Two neurons in the input layer represent the distance and time while the two neurons in the output layer represent the real and imaginary parts of NLSE solutions respectively. Both inputs and outputs of the model should be normalized and denormalized before entering and after leaving the fiber transmission model respectively. These operations are necessary since they can effectively map the actual inputs to the range where the model can be trained at the best efficiency. Since normalization and denormalization operations vary for different tasks, their detailed formats will be introduced in section B.

The design of loss functions is crucial since it can affect both convergence speed and final accuracy of the model when training. Since this model's essence is to approximate the solutions of NLSE, at least the equation, referred as NLSE and the initial conditions, referred as pulses or signals before transmission should be included in the loss function so that the whole model can be converged to obey both at the same time [11-12]. One thing needs to be noticed is that all values and loss functions in this principle-driven PINN based fiber model can only deal with discrete data which means continuous signals and solution planes of NLSE are all ought to be discretized and sampled in advance. Since detailed forms of equations and initial conditions are closely related with tasks, they will be introduced in section B as well.

When it comes to the general training and testing configurations, widely-utilized optimization methods based on gradient descend should be adopted on the condition that the gradients of loss functions towards networks' weights are analytical. In this principle-driven PINN based model, since its loss functions contain both partial differential equations and initial conditions, it is necessary to solve partial derivatives of network's outputs with respect to its inputs so as to finally obtain the values of the loss functions. Auto-differentiation tools which had been inserted in the Tensorflow architecture can solve this problem and thus will be adopted to obtain the gradients of loss functions in all three tasks conveniently[11-14]. For most neural networks, original or modified stochastic gradient descend algorithms are adopted for optimization. These methods, starting from a set of the randomly selected trainable parameters, force the networks converge to the desired states by utilizing the information of the first-order derivatives of loss functions [15-17]. Though alleviating the requirements for storing and computing, these algorithms may cause relatively slow and fluctuational convergence especially at the last several epochs of training since the first-order derivatives provide imperfect directions of gradient descending especially during this training stage. In order to overcome these drawbacks, BFGS [11,19] algorithms originated from Newton-Cortex iterative algorithm [20] are adopted to finely tune to model as to further the minimize the loss function values. Compared with ADAM optimizer, BFGS algorithms focus on the second-order Hessian matrices of the loss functions which can provide more accurate direction of gradients' descending though it may increase the requirements of computation and storage for gradients. In order to address this problem, L-BFGS algorithm [11,19] was invented. Since this method adopts Hessian matrices approximations instead of precise values, it can keep great balance between the accuracy of the gradient descending directions and the requirements of of computation and storage. In the original PINN networks, the authors provided both ADAM and L-BFGS optimizers in the training scenarios of PINN [11,18]. In this paper, we utilize both in sequence. In the early few training stages, ADAM optimizer is utilized in order to reach the minimum point at a relatively high speed, L-BFGS optimizer is then taken into use to finely tune the model to attain its best performances at latter stages of model training.

*B. Detailed model configurations for tasks*

As is mentioned in section A, many detailed configurations containing data normalization, loss functions etc are closely related with tasks. This section will first introduce the physical backgrounds and mechanisms of all three tasks and then connects them to the detailed configurations of the principle-driven PINN based fiber transmission model. All fiber parameters for three different tasks are listed in Table I. In this paper, all three tasks share the same core structure of PINN fiber model but with different hyper-parameters. However, since different tasks are based on slightly different formats of NLSE and signal shapes before transmission, each task adopts different normalization operations and loss functions.

*Task I: Pulse Evolution*

The pulse evolution task aims to test the model performances on predicting the changes of single or multiple pulses during fiber propagation. For this task, the inputs of the principle-driven PINN based fiber transmission model are time $T$ under time-retarded frame and propagation distance $z$ while the outputs are real and imaginary parts of pulse electrical field distribution values $\psi(T, z)$. However, if $(T, z)$ is input directly without normalization, it will be of high probability of causing gradient explosion and diminish phenomena, resulting in bad training results or even training failures. Thus, necessary frame transform in order to normalize both inputs and de-normalize outputs must be taken into use to avoid such situation. By adopting frame transform described in (3), the original NLSE [1,11-12] from fiber optics describing pulse evolutions can be normalized with respect to the normalized time retarded frame $t$ and distance frame $\zeta$ as

$$iA_1 \frac{\partial U}{\partial \zeta} + i\kappa_1 A_2 U + \kappa_1 A_3 \frac{\partial^2 U}{\kappa_2^2 \partial t^2} + i\kappa_1 A_4 \frac{\partial^3 U}{\kappa_2^3 \partial t^3}$$
$$+ \kappa_1 A_5 |U|^2 U + i\kappa_1 A_6 \frac{\partial |U|^2 U}{\kappa_2 \partial t} + \kappa_1 A_7 U \frac{\partial |U|^2}{\kappa_2 \partial t} = 0 \quad (1)$$

where parameters from $A_1$ to $A_7$ refer as

$$A_1 = 1 \quad A_2 = \frac{\alpha L_D}{2} \quad A_3 = -\frac{sign(\beta_2)}{2} \quad A_4 = -\frac{\beta_3 L_D}{6T_0^3} \quad A_5 = \frac{L_D}{L_{NL}}$$

$$A_6 = \frac{\gamma P_0 L_D}{\omega_0 T_0} \quad A_7 = -\frac{\gamma P_0 L_D \tau}{T_0} \quad L_D = \frac{T_0^2}{|\beta_2|} \quad L_{NL} = \frac{1}{\gamma P_0} \quad (2)$$

and the corresponding frame transform is

$$z = L_{max}\zeta = \frac{L_{max}}{L_D} L_D \zeta = \kappa_1 L_D \zeta$$

$$T = T_{max} t = \frac{T_{max}}{T_0} T_0 t = \kappa_2 T_0 t \quad \psi = \sqrt{P_0} U \quad (3)$$

where $U=U(t,\zeta)$ is the solution of (1) whose diagram is a complex surface while $\psi=\psi(T,z)$ represents the solution of the original NLSE describing pulse evolution. $T_0$ represents the full width of which the corresponding value equals 1/e of the pulse peak. $\alpha$ represents power attenuation per distance. $\beta_2$ and $\beta_3$ is the second and third order propagation constant which reflects dispersion and high order dispersion effects in the fiber respectively. $\gamma$ represents nonlinear coefficient. $\omega_0$ describes the central angular frequency of the light pulse. $\tau$ is related with the delayed Raman response. In conclusion, when using this principle-driven PINN based fiber model, (3) is firstly utilized to normalize the original physical frames into the model's standard inputs. In the end, outputs of PINN based model can be transformed back into the original physical values by using (3). This process of normalization and denormalization is vividly depicted in Fig.1(b) as well.

As for the loss function $J$ for this task, it should contain the normalized and discretized NLSE shown in (1) which describes pulse evolution and sampled normalized pulse profiles. It can be described as

$$J = J_1 + J_2$$
$$= \frac{1}{N_{ini}} \sum_{k=1}^{N_{ini}} \left| U_P^k(0, t^k) - U^k(0, t^k) \right|^2$$
$$+ \frac{1}{N_P} \sum_{i=1}^{N_P} \left| \begin{array}{l} iA_1 \frac{\partial U_P^i}{\partial \zeta} + i\kappa_1 A_2 U_P^i + \kappa_1 A_3 \frac{\partial^2 U_P^i}{\kappa_2^2 \partial t^2} + i\kappa_1 A_4 \frac{\partial^3 U_P^i}{\kappa_2^3 \partial t^3} \\ + \kappa_1 A_5 \left| U_P^i \right|^2 U_P^i + i\kappa_1 A_6 \frac{\partial \left| U_P^i \right|^2 U_P^i}{\kappa_2 \partial t} + \kappa_1 A_7 U_P^i \frac{\partial \left| U_P^i \right|^2}{\kappa_2 \partial t} \end{array} \right|^2 \quad (4)$$

where $U_P$ represents complex solution of (1) which are originated from the outputs of the principle-driven PINN based fiber transmission model while $N_{ini}$ and $N_P$ represents the number of the sampling points of model's inputs and outputs respectively. In (4), $J_1$ measures the distance between the initial profile of the outputs and the initial pulse profile before transmission while $J_2$ the extent of wellness about how the outputs from the PINN based fiber transmission model fits NLSE and initial conditions.. From (4), it also can also be seen that automatic neural network differentiation tool which is proposed previously in many papers [13-14] and inserted in the architecture of Tensorflow is necessary to use so that gradients of loss function can be obtained.

*Task II: Signal transmission*

Unlike the first task focusing on pulses evolution, this task aims at testing the signal prediction performances of the principle-driven PINN based fiber transmission model on signals with random code patterns. Since stochastic code patterns exist in this task, it will be harder for the model to accomplish than that of the pulse evolution task.

When conducting the task of signal transmission, the initial conditions are different since in this task, irregular time-domain waveforms carrying transmitted information can not be described analytically like regular pulses even in the simplest cases. Parameters which are adopted to measure pulses' width and maximum power are no longer suitable to measure the properties of transmitted signals. Therefore, it is inappropriate and meaningless to use former parameters like $T_0$ to describe and normalize both initial signals before fiber transmission and NLSE. Instead, symbol duration $T_s$ which is the reciprocal of the symbol rate is introduced to describe time-domain width for each duration of the one transmitted symbol. Thus, the original NLSE [1] can be normalized as

$$iB_1 \frac{\partial s}{\partial \zeta} + i\chi_1 B_2 s + \chi_1 B_3 \frac{\partial^2 s}{\chi_2^2 \partial t^2} + i\chi_1 B_4 \frac{\partial^3 s}{\chi_2^3 \partial t^3}$$
$$+ \chi_1 B_5 |s|^2 s + i\chi_1 B_6 \frac{\partial |s|^2 s}{\chi_2 \partial t} + \kappa_1 B_7 s \frac{\partial |s|^2}{\chi_2 \partial t} = 0 \quad (5)$$

where parameters from $B_1$ to $B_7$ equals

$$B_1 = 1 \quad B_2 = \frac{\alpha L_D}{2} \quad B_3 = -\frac{sign(\beta_2)}{2} \quad B_4 = -\frac{\beta_3 L_D}{6T_s^3} \quad B_5 = \frac{L_D}{L_{NL}}$$
$$B_6 = \frac{\gamma P_{max} L_D}{\omega_0 T_s} \quad B_7 = -\frac{\gamma P_{max} L_D \tau}{T_s} \quad L_D = \frac{T_B^2}{|\beta_2|} \quad L_{NL} = \frac{1}{\gamma P_{max}} \quad (6)$$

and the corresponding frame transform is defined as

$$z = L_{max}\zeta = \frac{L_{max}}{L_D} L_D \zeta = \chi_1 L_D \zeta$$
$$T = T_{max} t = \frac{T_{max}}{T_s} T_s t = \chi_2 T_s t \quad S = \sqrt{P_{max}} s \quad (7)$$

in which $P_{max}$ represents the maximum power of the transmitted modulated signal and $s = s(t,\zeta)$ represents the normalized signal with respect to normalized transmission distance $\zeta$ and normalized time-retarded frame $t$.

Loss function $J$ of this task can then be determined after taking into consideration of the normalized NLSE under the task of fiber transmission which measures the extent of how well the outputs from the principle-driven PINN based fiber transmission model satisfies NLSE and initial conditions. After sampling the initial signals profile before fiber transmission and the solution plane, the loss function $J$ of this task can be expressed as

$$J = J_1 + J_2$$
$$= \frac{1}{N_{ini}} \sum_{k=1}^{N_{ini}} \left| s_P^k(0, t^k) - s^k(0, t^k) \right|^2$$
$$+ \frac{1}{N_P} \sum_{i=1}^{N_P} \left| \begin{array}{l} iB_1 \frac{\partial s_P^i}{\partial \zeta} + i\chi_1 B_2 s_P^i + \chi_1 B_3 \frac{\partial^2 s_P^i}{\chi_2^2 \partial t^2} + i\chi_1 B_4 \frac{\partial^3 s_P^i}{\chi_2^3 \partial t^3} \\ + \chi_1 B_5 \left| s_P^i \right|^2 s_P^i + i\chi_1 B_6 \frac{\partial \left| s_P^i \right|^2 s_P^i}{\chi_2 \partial t} + \chi_1 B_7 s_P^i \frac{\partial \left| s_P^i \right|^2}{\chi_2 \partial t} \end{array} \right|^2 \quad (8)$$

The model will converge to learn to obtain the relations between signals before transmission and transmitted signals progressively over the training procedures which can minimize the value of loss function (8). After being trained appropriately, the principle-driven PINN based fiber transmission model will obey NLSE describing signal transmission and initial signals before fiber transmission at the same time.

*Task III: Fiber birefringence*

The last task, which is the most hardest, aims at testing the performance of the principle-driven PINN based fiber transmission model on solving fiber birefringence problems described by the polarized dimensional NLSE equations which can also be referred as Manakov Equations. This equations take polarization effect which is ignored in the previous two tasks into consideration. Fiber birefringence caused by asymmetric fiber intersection or refraction index difference is an important effect which is closely related with polarization in fiber optics. Stochastic, time-varying fiber birefringence effect is also regarded as polarization mode dispersion (PMD) in fiber optics which can degrade the quality of transmitted signals. In this paper, we only consider the prediction task of pulse evolution in fiber with constant, time-invariant birefringence effect for simplicity without lose generality.

Fixed fiber birefringence effect can be described precisely by Manakov Equation [1] which contains two NLSE equations describing pulse evolution in orthogonal axis-$x$ and $y$ in contrast with propagation axis $z$. This can be written as

$$\begin{cases} \dfrac{\partial \psi_x}{\partial z}+\dfrac{\alpha}{2}\psi_x+\beta_{1x}\dfrac{\partial \psi_x}{\partial T}+i\dfrac{\beta_2}{2}\dfrac{\partial^2 \psi_x}{\partial T^2}-i\gamma\left(|\psi_x|^2+\dfrac{2}{3}|\psi_y|^2\right)\psi_x=0 \\ \dfrac{\partial \psi_y}{\partial z}+\dfrac{\alpha}{2}\psi_y+\beta_{1y}\dfrac{\partial \psi_y}{\partial T}+i\dfrac{\beta_2}{2}\dfrac{\partial^2 \psi_y}{\partial T^2}-i\gamma\left(|\psi_y|^2+\dfrac{2}{3}|\psi_x|^2\right)\psi_y=0 \end{cases} \quad (9)$$

where $\psi_x$ and $\psi_y$ represents pulse projection component in the polarization direction towards $x$-axis and $y$-axis respectively. $\beta_{1x}$ and $\beta_{1y}$ represents the first-order propagation constant of $x$-direction polarization and $y$-direction polarization in the fiber respectively.

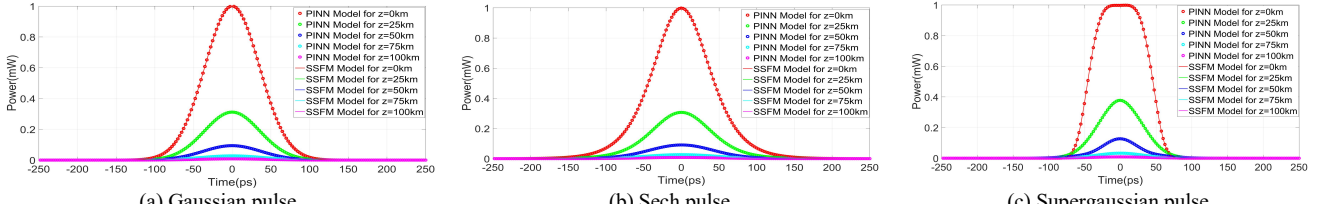

Fig.2. Performances of PINN-based fiber model by single pulse as the first case of pulse evolution task

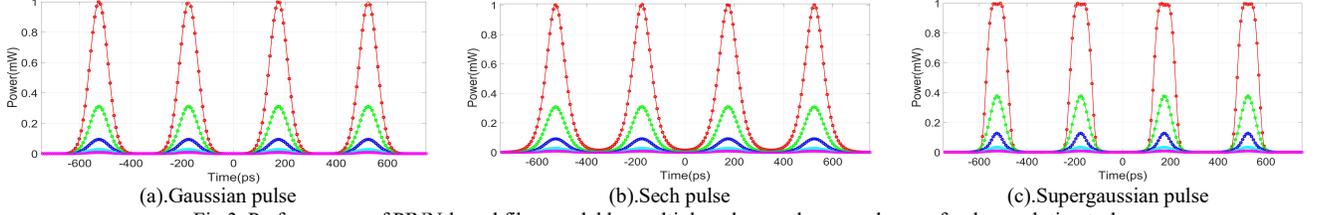

Fig.3. Performances of PINN-based fiber model by multiple pulses as the second case of pulse evolution task

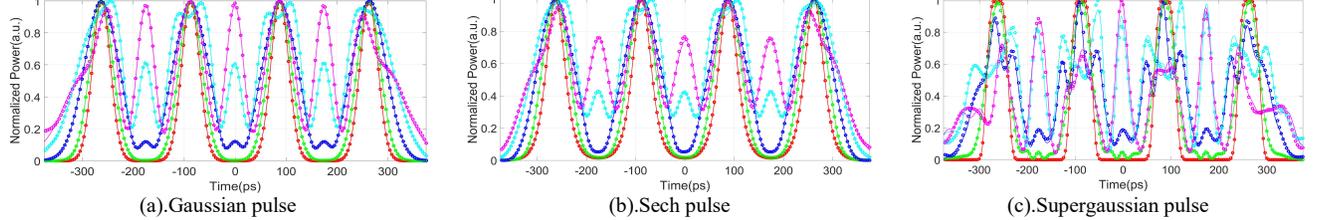

Fig.4. Performances of PINN-based fiber model by multiple pulses with shorter $T_0$ as the third case of pulse evolution task

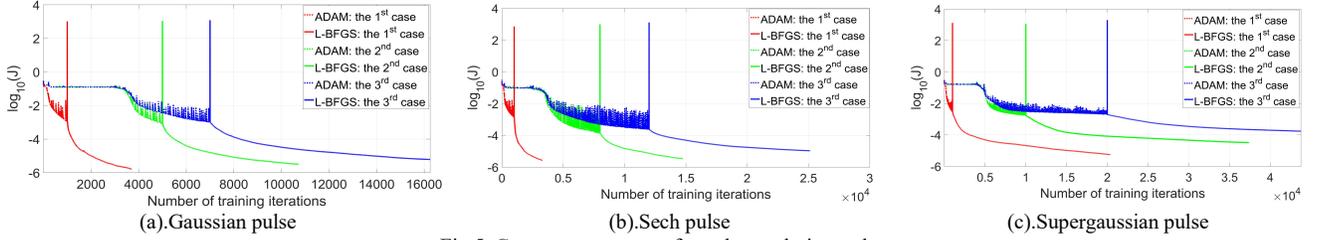

Fig.5. Convergence curves for pulse evolution task

Like normalization operations in tasks of pulse evolution and signal transmission, (9) also needs to be normalized in order to apply into PINN based fiber model. By using the similar normalized method, it can be written as [1]

$$\begin{cases} A_1\dfrac{\partial U_x}{\partial \zeta}+A_2\kappa_1 U_x+A_3\kappa_1\dfrac{\partial U_x}{\kappa_2\partial t}+iA_4\kappa_1\dfrac{\partial^2 U_x}{\kappa_2^2\partial t^2}+iA_5\kappa_1\left(P_{0x}|U_x|^2+\dfrac{2}{3}P_{0y}|U_y|^2\right)U_x=0 \\ A_1\dfrac{\partial U_y}{\partial \zeta}+A_2\kappa_1 U_y-A_3\kappa_1\dfrac{\partial U_y}{\kappa_2\partial t}+iA_4\kappa_1\dfrac{\partial^2 U_y}{\kappa_2^2\partial t^2}+iA_5\kappa_1\left(P_{0y}|U_y|^2+\dfrac{2}{3}P_{0x}|U_x|^2\right)U_y=0 \end{cases} \quad (10)$$

where parameters from $A_1$ to $A_5$ equals

$$A_1=1 \quad A_2=\dfrac{\alpha L_D}{2} \quad A_3=\dfrac{\Delta\beta_1 L_D}{2T_0} \quad A_4=-\dfrac{sign(\beta_2)}{2} \quad A_5=\gamma L_D \quad L_D=\dfrac{T_0^2}{|\beta_2|} \quad (11)$$

$$z=\kappa_1 L_D\zeta \quad T=\kappa_2 T_0 t \quad \psi_x=\sqrt{P_{0x}}U_x \quad \psi_y=\sqrt{P_{0y}}U_y \quad (12)$$

in which $P_{0x}$ and $P_{0y}$ represents the maximum power of the projection component in the two polarization directions.

The loss function $J$ of this task should thus contain normalized desretized Manakov Equations and sampled pulse profiles before fiber transmission in $x$ and $y$ direction which can be illustrated correspondingly as

$$J=\dfrac{1}{N_{ini}}\sum_{k=1}^{N_{ini}}\left|U_{xp}^k(0,t^k)-U_x^k(0,t^k)\right|^2$$
$$+\dfrac{1}{N_P}\sum_{i=1}^{N_P}\left|\begin{array}{l}A_1\dfrac{\partial U_{xP}^i}{\partial \zeta}+\kappa_1 A_2 U_{xP}^i+A_3\dfrac{\partial U_{xP}^i}{\kappa_2\partial t}+i\kappa_1 A_4\dfrac{\partial^2 U_{xP}^i}{\kappa_2^2\partial t^2}\\ +i\kappa_1 A_5\left(P_{0x}|U_{xP}^i|^2+\dfrac{2}{3}P_{0y}|U_{yP}^i|^2\right)U_{xP}^i\end{array}\right|^2$$
$$+\dfrac{1}{N_{ini}}\sum_{k=1}^{N_{ini}}\left|U_{yp}^k(0,t^k)-U_y^k(0,t^k)\right|^2$$
$$+\dfrac{1}{N_P}\sum_{i=1}^{N_P}\left|\begin{array}{l}A_1\dfrac{\partial U_{yP}^i}{\partial \zeta}+\kappa_1 A_2 U_{yP}^i-A_3\dfrac{\partial U_{yP}^i}{\kappa_2\partial t}+i\kappa_1 A_4\dfrac{\partial^2 U_{yP}^i}{\kappa_2^2\partial t^2}\\ +i\kappa_1 A_5\left(P_{0y}|U_{yP}^i|^2+\dfrac{2}{3}P_{0x}|U_{xP}^i|^2\right)U_{yP}^i\end{array}\right|^2 \quad (13)$$

III. SIMULATION RESULTS AND DISCUSSIONS

In this part, the simulation results of the principle-driven PINN based fiber model on three tasks will be shown and its performances will be analyzed in detail. Different diagrams of model predictions including time-domain waveforms, eye diagrams etc. will be depicted based on the specifications and characteristics of each task in order to better reflect and measure the model's accuracy. For most cases, waveforms at different transmission distances are chosen to be shown in order to better depict the details of transmitted pulses/signals. Calculated signals after transmission from SSFM based fiber transmission model are utilized as standard references for comparisons. Further discussions on model's performances will be developed after results analysis.

*A. Simulation results*

*Task I: Pulse evolution*

As for the test results of pulse evolution, important parameter settings of fiber for this task are illustrated in the second column of Table I. Note that the value of dispersion slope is set to be relatively large in order to test the model's performances in predicting pulses propagating through fibers with relatively intense high-order dispersion effect. In total three cases of configurations of pulses before fiber transmission have been designed to test the performances of the principle-driven PINN based fiber transmission model. Pulses of three different shapes including Gaussian pulse, Sech pulse and second-order Supergaussian pulse are used in all three case. The first case only deals with single pulse which is the simplest among all. $T_0$ for all pulses is 50ps and the peak power $P_0$ for all pulses is 1mW. Fig.2(a)-(c) depicts the outputs from PINN based model and SSFM based fiber model. Five different colors-red, green, blue, cyan and pink are used to describe the pulse's shapes at 0, 25, 50, 75 and 100km while data dots and solid lines are utilized to represent outputs from the principle-driven PINN based fiber model and SSFM based fiber model respectively. In the remaining two cases, four pulses as a whole are adopted as the initial condition without losing generality as to test the performance of PINN based fiber model on predicting rather sophisticated pulse shapes during fiber propagation. In the second case, the interval time between each two adjacent pulses is set to be 0.35ns. The performances of PINN based model in this case on the contrast with SSFM based model is shown in Fig.3(a)-(c). In the last case, in order to show the more intense pulse interference during propagation, both interval time between two neighbouring pulses and $T_0$ of the pulse is set to be the half. Corresponding results of this case are shown in Fig.4(a)-(c) where high-order dispersion phenomena can be clearly seen in this case since this effect is much more notable when $T_0$ of pulse is small as the pulse evolves asymmetrically through fiber.

Generally speaking, as the complexity of initial conditions increases as either more sophisticated shapes or more numbers of pulses are adopted as initial conditions, the prediction accuracy of PINN based fiber model decreases. This can be concluded by comparing pulses with different waveforms and width from Fig.2-Fig.4. Flaws of the output predictions of the principle-driven PINN based fiber model are mostly shown at the peaks of the pulses from model's outputs which slightly oscillations exist while it should be flat both in theory and outputs from SSFM based fiber model.

Convergence performances are illustrated in detail in Fig. 5(a)-(c). The obvious breaking point in the convergence curve divides the training procedure into two parts. In the first part whose convergence curves are depicted as the dashed lines, the ADAM optimizer is firstly adopted to train the network at a relatively high convergence rate. In the second stage, the L-BFGS optimizer is used to further train the network, of which the convergence curves are plotted by solid line. This phenomenon confirms the optimizer configurations illustrated in the section of model structures and training configurations. The overall decrease in the slopes of convergence curves indicates that the difficulties of both initial pulse's shapes and fiber effects contained can indeed affect the performances of convergence as well. When the initial pulse's profile changes from Gaussian pulse to Supergaussian pulse, and when the number of initial pulses before fiber transmission increases, not only will the convergence rate slow down but also final accuracy will decrease.

*Task II: Signal transmission*

As for the second signal transmission task, important parameters of the fiber can be found from the third column of Table I. The transmission rate is set to be 10GBaud, 20GBaud and 40GBaud whose corresponding $T_s$ equals 100ps, 50ps and 25ps. Fig 6(a)-(c) shows the panorama view of time-domain signals changes during transmission which indicates that this principle-driven PINN based fiber transmission model can show the whole view on signal changes during fiber transmission for only one computation.

As can be seen from the panorama view, when the symbol rate increases which indicates shorter symbol period, the signals experience more severe distortions due to fiber dispersion effect, resulting in worse inter-symbol interference (ISI) at a rather earlier transmission distance. As for the principle-driven PINN based fiber model, higher symbol rate implies more difficulties in predictions since the solutions of NLSE vary more frequently and fiercely with respect to the standard normalized solution planes of time and distance.

In order to better describe the quality of signal transmitted, eye diagrams of the signals at different transmission distances equaling 0, 25, 50 and 100km from both SSFM and PINN based fiber models are depicted in Fig.7 as well. By inspecting these eye diagrams from the two models, not only signal transmission quality but also signal prediction performances of the principle-driven PINN based fiber model can be indicated clearly.

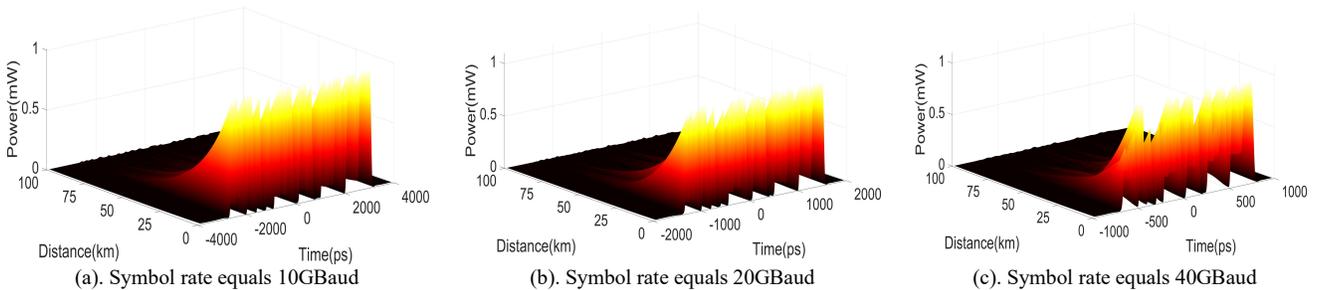

(a). Symbol rate equals 10GBaud      (b). Symbol rate equals 20GBaud      (c). Symbol rate equals 40GBaud

Fig.6. Time domain signals for signal transmission task

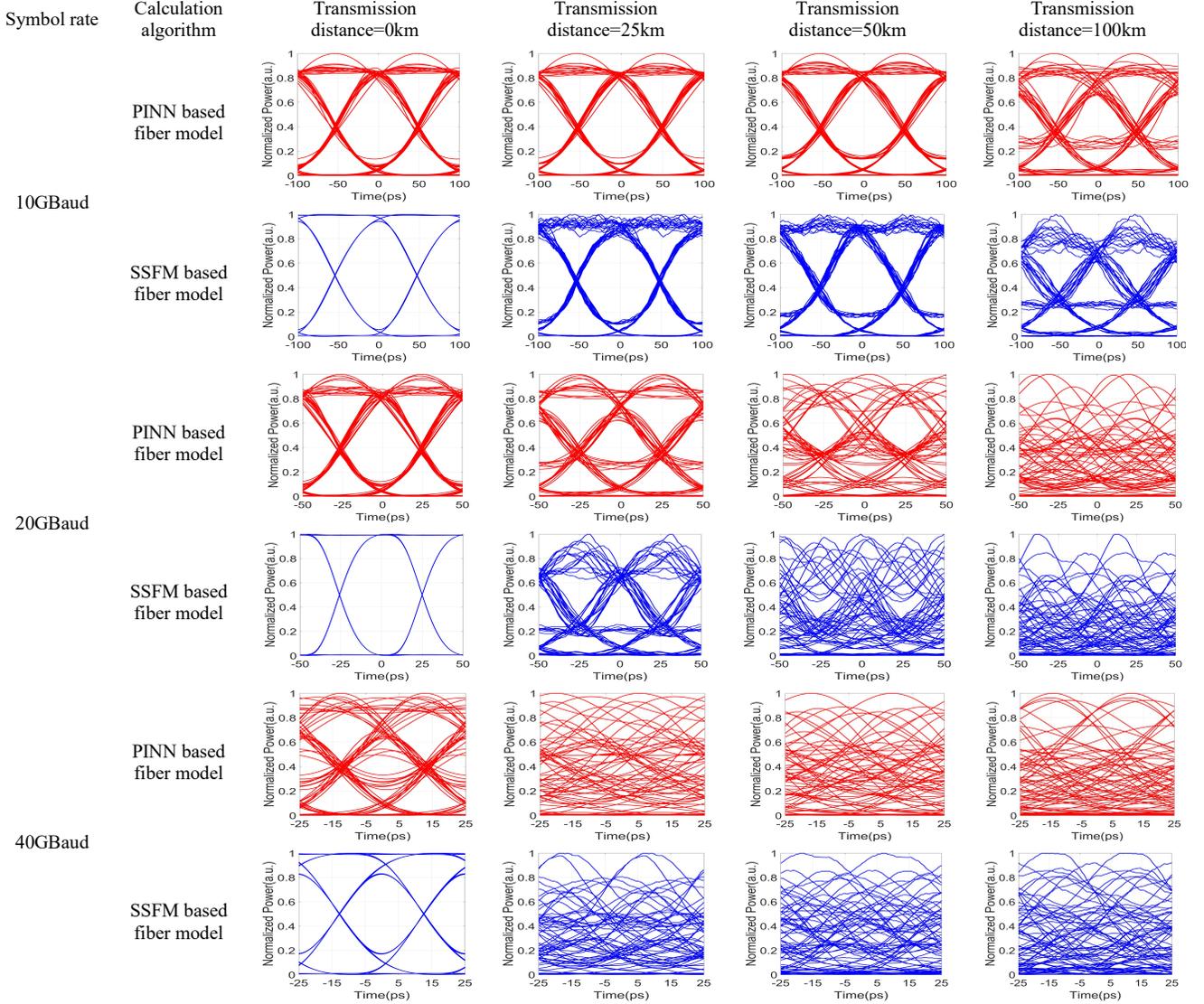

Fig. 7. Eye diagram for signal transmission task

When the symbol rate equals 10GBaud, the structure of eye diagrams is relatively clear for all transmission distances which indicates that the signal transmission quality under this circumstance is relatively better for detection. In contrast with the eye diagram from signals with the same transmission distance but different symbol rate, the eye-shaped structure becomes blur for higher symbol rate. This clearly indicates that the quality of the signal transmission degrades for higher transmission symbol rate. By comparing eye diagram from PINN based fiber model and SSFM based fiber model, differences can be clearly seen. For low symbol rate, eye diagrams from these two different models are relatively much more similar except few differences when transmission distance is longer. This conclusion can be vividly drawn by comparing subfigures from the first and second row. The differences may become much more obvious as symbol rate increases. Under the circumstance where symbol rate equals 40GBaud, even when transmission distance equals 0km, the shape of eye diagram of the principle-driven PINN based fiber transmission model shows relatively more notable differences in the thickness of lines in eye diagram, not to mention those differences of eye diagrams from two models in longer transmission distances. This indicates that the prediction errors of PINN based fiber model increase as symbol rate increases which causes relatively more fierce ISI.

*Task III: Fiber birefringence*

In order to numerically demonstrate the prediction ability of the principle-driven PINN based fiber model for fiber birefringence effect, pulses of three different shapes-standard Gaussian, Sech and Super-Gaussian are adopted as well without losing generality. These linear-polarized light pulses are assumed to induce into fiber at the angle of $\pi/4$ towards $x$-axis and then propagates toward $z$ axis in the fiber without losing generality. All important parameter settings of the fiber simulated in this task are clearly shown in the last column of Table I. The simulation results are shown in Fig.8. For all figures expect Fig.8(c), (g) and (j), Red, green, blue, cyan and pink are used to represent pulses' shapes at 0, 25, 50, 75, 100km respectively while for the remaining three figures, they are used to represent pulses at 0, 5, 10, 15, 20km since under this circumstance where the difference between first-order propagation constant in *x* and *y* axis $\Delta\beta_1$ equals $2\times10^{-14}$ s/m and the fiber birefringence effect is so intense that pulses will soon split after the transmission of over 20km. In order to testify the accuracy of the principle-driven PINN-based fiber model, data dots are utilized to represent its results while solid lines are used to show results from SSFM-based model.

Due to the fiber birefringence effect, the pulse transmitted gradually presents the phenomenon of broadening and splitting. By comparing results of different $\Delta\beta_1$, it can be concluded that with the increase of $\Delta\beta_1$, pulses' broadening and splitting tend to occur earlier and more fiercely. In all three cases, the principle-driven PINN-based fiber model can obtain the relatively high precise results as all data dots fall at the solid lines in Fig.8. Due to the induced angle, both component of *x* and *y* direction obtain the same amount of pulse power which can be clearly seen especially when pulse splitting occurs. When the light pulse induces at a different angle, the power of these two components may differ, resulting in asymmetric shape when pulse broadening and splitting.

into considerations by designing different loss functions for different tasks. For other tasks in fiber optics and telecommunications which are not mentioned in this paper, normalization operations, loss functions designs, network training and testing scenarios can be conducted similarly.

Both model structures and hyper-parameters should be dynamically adjusted with respect to different tasks. Large scale networks can take advantages in its rather higher order trainable parameter space so that better performances can of high probability of being obtained on sophisticated tasks though storage requirements are higher. As a consequence, it is highly recommended that the appropriate network scale should be well determined so as to better balance the signal prediction performances with the storage requirements.

In most neural networks or data-driven models which adopt large amount of labeled data for training and testing, when applying these networks on rather easier tasks or smaller training data scale, overfitting phenomenon especially over fiber length may have high probability of occurring since these models can only learn the rules of fiber transmission from the distribution of data inputs and labels. On the contrast, this distance overfitting problem can be effectively avoided in our model since physical principles instead of data labels are introduced in loss functions to help converge the model to the

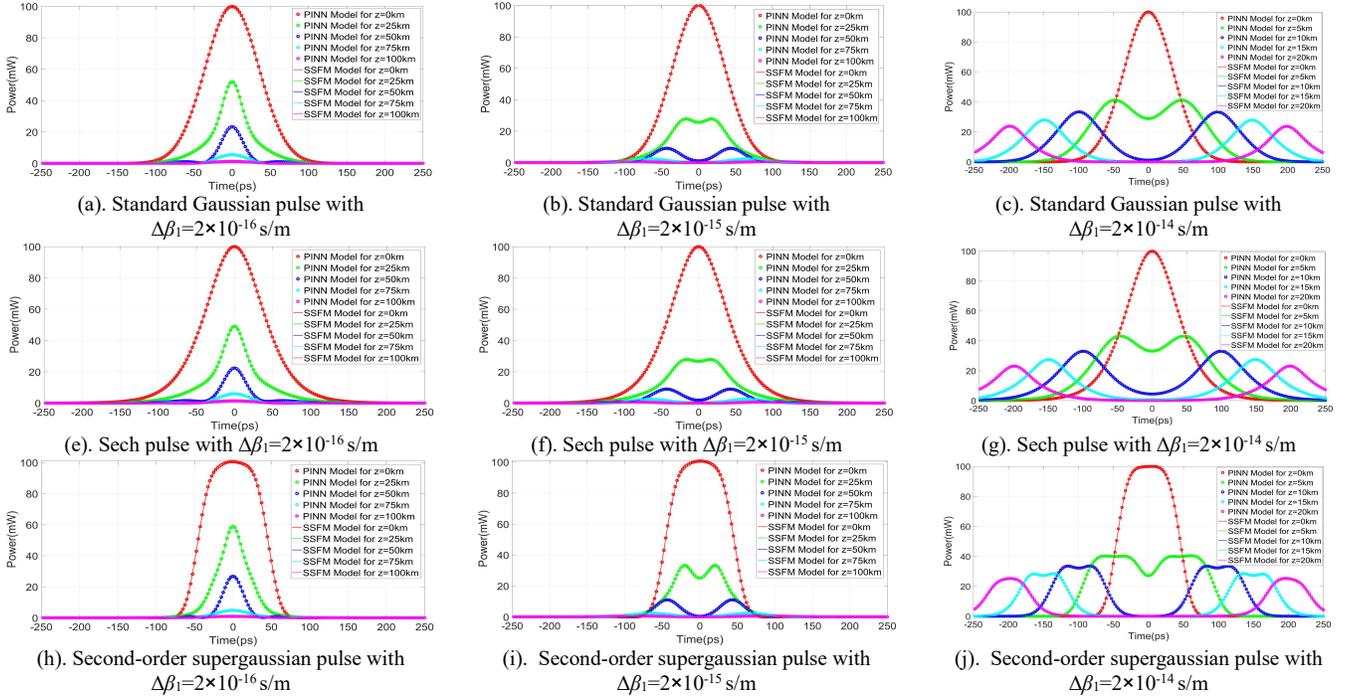

(a). Standard Gaussian pulse with $\Delta\beta_1=2\times10^{-16}$ s/m

(b). Standard Gaussian pulse with $\Delta\beta_1=2\times10^{-15}$ s/m

(c). Standard Gaussian pulse with $\Delta\beta_1=2\times10^{-14}$ s/m

(e). Sech pulse with $\Delta\beta_1=2\times10^{-16}$ s/m

(f). Sech pulse with $\Delta\beta_1=2\times10^{-15}$ s/m

(g). Sech pulse with $\Delta\beta_1=2\times10^{-14}$ s/m

(h). Second-order supergaussian pulse with $\Delta\beta_1=2\times10^{-16}$ s/m

(i). Second-order supergaussian pulse with $\Delta\beta_1=2\times10^{-15}$ s/m

(j). Second-order supergaussian pulse with $\Delta\beta_1=2\times10^{-14}$ s/m

Fig.8. Simulation results of fiber refringence effect

### B. Further discussions

The above numerically demonstrations indicate that the principle-driven PINN based fiber model can have relatively good signal prediction ability in pulse evolution, signal transmission and fiber birefringence. Unlike either conventional SSFM based fiber transmission models or data-driven fiber transmission models, this model operates as a NLSE equations solver based on fully-connected neural network structures with different initial pulses/signals conditions for different tasks. Physical principles are taken desired state. This guarantees that this model's outputs are independent from the distribution of training data.

More attention needs to be paid on further improving the performances of PINN based fiber model on predicting more complicated tasks such as modeling fibers with more intense non-linearity, conducting predictions on signals with high-order modulation formats. Besides, as for the current numerical demonstrations, some of the factors such as noise, stochastic PMD are neglected or partially simplified while they may still play important roles in degrading the quality of transmitted signals. These factors will be taken into considerations in further research as to make the whole principle-driven PINN based fiber model more universal.

## IV. Conclusion

In this paper, we apply PINN into fields of fiber transmission model by modifying the original PINN proposed by Maziar Raissi et al mainly through considering more effects, normalizing frame and equations, appropriately designing loss functions and connecting with actual physical mechanisms including pulses before transmission and NLSE. Since this model takes the fiber transmission as NLSE solving problems with initial pulses/signals conditions, no labeled data are needed while training. Instead of operating as black boxes like data-driven models, this principle-driven model takes NLSE as the physical meaning and initial condition as data information into loss functions and can progressively learn the fiber propagation rules during training procedures.

Three main advantages can thus be obtained once this model is adopted. Firstly, compared with conventional SSFM based fiber model, the number of fiber sections is no longer needed to be determined so that the risk of wrong results due to misconfigurations can be effectively avoided since the model parameters will be the same for different transmission distances once trained. Secondly, compared with data-driven models, no signals after transmission are needed to be calculated by SSFM in advance before model training since this model converges in maximize the extent of wellness about how the outputs from the principle-driven PINN based fiber transmission model fits NLSE and initial conditions. Therefore, time consumption should be low compared with data-driven fiber models since SSFM is no longer needed for data preparations before training. Besides, since the principle-driven model learns transmission rules through physical principles rather than data, there will be no worries of the model's dependence on the distribution of training data. Thirdly, this model can provide the views on how pulses or signals change all along the whole processes of fiber transmission for one computation while both SSFM based and data-driven models may need to compute multiple times. This advantage may provide great convenience for fiber optics researching.

Through numerical demonstration, results indicate that the principle-driven PINN based fiber model can have the potential of becoming a new effective way in accomplishing the tasks of predicting pulse evolution, signal transmission and fiber birefringence effect. Further conclusion can be drawn from the procedures of training and testing that different training hyper-parameters can result in different prediction performances. These hyper-parameters such as number of layers, neurons, sampling points in solution domain and maximum iterations for training etc. barely exist analytical rules for configurations. However, more complicated fiber transmission circumstances tend to need more complicated model structures and hyper-parameters in order to cater the need for signal predictions.

## Acknowledgment

This work was supported by the National Key Research and Development Program of China under Contract 2019YFB1803501, the National Natural Science Foundation of China under Contract 61771284 and a grant from the Institute for Guo Qiang Tsinghua University.